\documentstyle[11pt,epsf]{article}

\def\appendix{\setcounter{section}{0}
\def\thesection{Appendix \Alph{section}}
\def\theequation{\Alph{section}.\arabic{equation}}}

\newfont{\subsub}{cmr6}

\newcounter{szk}

\begin{document}
\title{The uniqueness of company size distribution function \\
from tent-shaped growth rate distribution}
\author{
Atushi Ishikawa
\\
Kanazawa Gakuin University, Kanazawa 920-1392, Japan
}
\date{}
\maketitle

\begin{abstract}
\indent
We report the proof that
the extension of Gibrat's law in the middle scale region is unique and 
the probability distribution function (pdf) is also uniquely derived from
the extended Gibrat's law and
the law of detailed balance.
In the proof, two approximations are employed.
The pdf of growth rate is described as tent-shaped exponential functions
and the value of the origin of the growth rate distribution is constant.
These approximations are confirmed in profits data of Japanese companies
2003 and 2004.
The resultant profits pdf fits with the empirical data with high accuracy.
This guarantees the validity of the approximations.
\end{abstract}
\begin{flushleft}
\end{flushleft}

In the large scale region of 
wealth, assets, sales, profits, income, the number of employees and etc ($x$),
the cumulative probability distribution function (pdf) $P(> x)$ obeys a power-law
\begin{eqnarray}
    P(> x) \propto x^{-\mu}~~~~{\rm for }~~~~x > x_0~,
    \label{Pareto}
\end{eqnarray}
where $x_0$ is a certain threshold.
This power-law is called Pareto's law~\cite{Pareto},
which has been well investigated by using various models.

A few years ago,
Fujiwara et al.~\cite{FSAKA} show that
Pareto's law 
is derived from 
Gibrat's law~\cite{Gibrat} and
the law of detailed balance
by using no model. 
Gibrat's law is observed in the large scale region.
It argues that
the conditional pdf of growth rate $Q(R|x_1)$ 
is independent of the initial value $x_1$: 
\begin{eqnarray}
    Q(R|x_1) = Q(R)~.
    \label{Gibrat}
\end{eqnarray}
Here growth rate $R$ is defined as $R = x_2/x_1$ and
$Q(R|x_1)$ is defined by using the pdf $P(x_1)$
and the joint pdf $P_{1 R}(x_1, R)$ as 
$Q(R|x_1) = P_{1 R}(x_1, R)/P(x_1)$.
The detailed balance observed in a stable economy
is time-reversal symmetry ($x_1 \leftrightarrow x_2$):
\begin{eqnarray}
    P_{1 2}(x_1, x_2) = P_{1 2}(x_2, x_1)~.
    \label{Detailed balance}
\end{eqnarray}
Here $x_1$ and $x_2$ are  two successive wealth, assets, sales, profits, income, the number of employees, etc. 
and $P_{1 2}(x_1, x_2)$ is the joint pdf.

This derivation is significant for the understanding of the mechanism of Pareto's law,
however, it is valid only in the large scale region
where Gibrat's law (\ref{Gibrat}) holds.
It is well known that Pareto's law is not observed 
below the threshold $x_0$~\cite{Gibrat, Badger}.
The reason is thought to be the breakdown of Gibrat's law~\cite{FSAKA}, \cite{Stanley1}--\cite{Aoyama}.

In Ref.~\cite{Ishikawa2}, Gibrat's law is extended in the middle scale region
by employing profits data of Japanese companies in 2002 and 2003.
We approximate the conditional pdf of profits growth rate 
as so-called tent-shaped exponential functions
\begin{eqnarray}
    Q(R|x_1)&=&d(x_1)~R^{-t_{+}(x_1)-1}~~~~~{\rm for}~~R > 1~,
    \label{tent-shaped1}\\
    Q(R|x_1)&=&d(x_1)~R^{+t_{-}(x_1)-1}~~~~~{\rm for}~~R < 1~.
    \label{tent-shaped2}
\end{eqnarray}
By measuring $t_{\pm}$ we have assumed the $x_1$ dependence to be
\begin{eqnarray}
    t_{\pm}(x_1)=t_{\pm}(x_0) \pm \alpha_{\pm}~\ln \frac{x_1}{x_0}~,
    \label{t}
\end{eqnarray}
and have estimated the parameters as~\cite{Ishikawa}
\begin{eqnarray}
    \alpha_{+} \sim \alpha_{-} &\sim& 0~~~~~~~~~{\rm for}~~x_1 > x_0~,
    \label{alphaH}\\
    \alpha_{+} \sim \alpha_{-} &\neq& 0~~~~~~~~~{\rm for}~~x_{{\rm min}} < x_1 < x_0~,
    \label{alphaM}\\
    t_{+}(x_0) - t_{-}(x_0) &\sim& \mu~.
    \label{mu}
\end{eqnarray}
From extended Gibrat's law (\ref{t}) -- (\ref{mu}) and
the detailed balance (\ref{Detailed balance}),
we have derived the pdf in the large 
and middle scale region uniformly as follows
\begin{eqnarray}
    P(x) = C x^{-\left(\mu+1\right)}~e^{-\alpha \ln^2 \frac{x}{x_0}}
    ~~~~~~~~~{\rm for}~~x > x_{{\rm min}}~,
    \label{HandM}
\end{eqnarray}
where $\alpha = \left(\alpha_{+}+\alpha_{-}\right)/2$.
This is confirmed in the empirical data.

In this study, we prove that the $x_1$ dependence of $t_{\pm}$ 
(\ref{t}) with $\alpha_+ = \alpha_-$ is unique
if the pdf of growth rate 
is approximated by tent-shaped exponential functions 
(\ref{tent-shaped1}), (\ref{tent-shaped2}).
This means, consequently, 
that the pdf in the large and middle scale region (\ref{HandM})
is also unique
if the $x_1$ dependence of $d(x_1)$ is negligible.
We confirm these approximations in profits data of Japanese companies 2003 ($x_1$) and 2004 ($x_2$)
\cite{TSR}
and show that the pdf (\ref{HandM}) fits with empirical data nicely.

\begin{figure}[htb]
 \begin{minipage}[htb]{0.49\textwidth}
  \epsfxsize = 1.0\textwidth
  \epsfbox{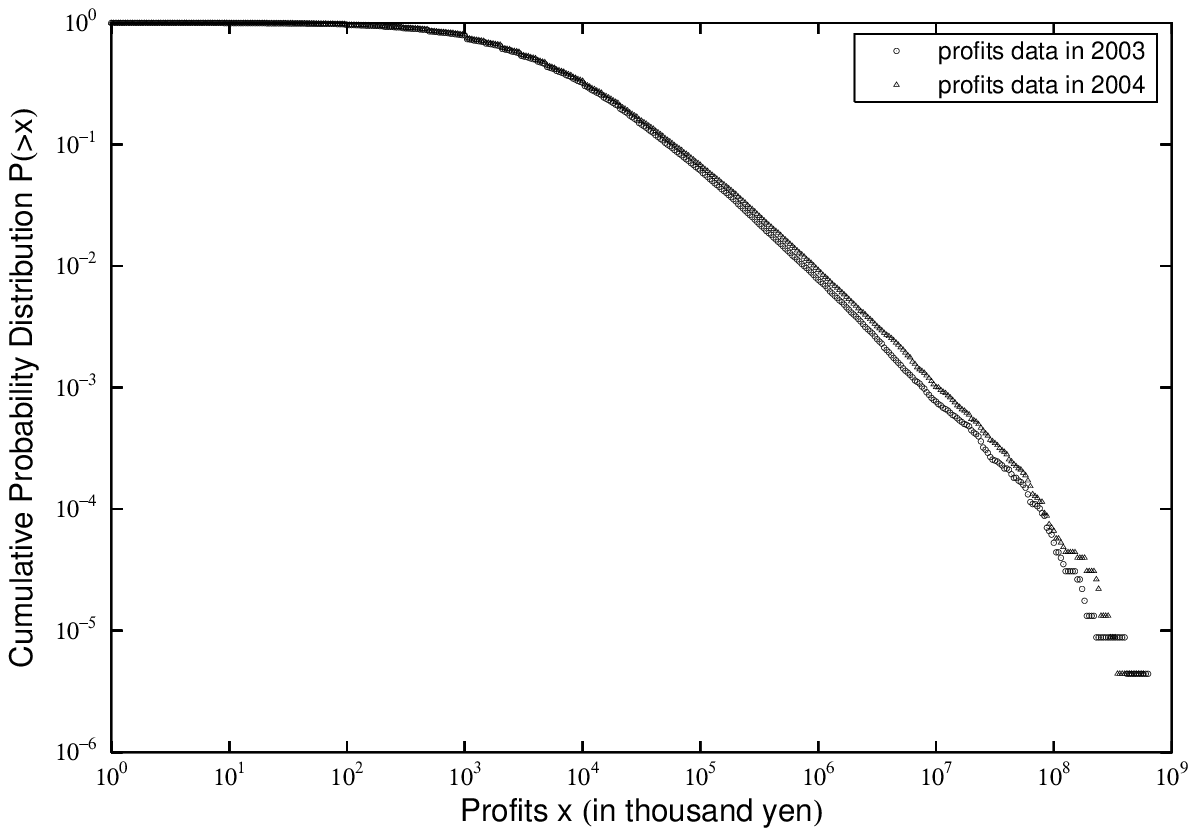}
 \caption{Cumulative probability distributions $P(> x_1)$ and $P(> x_2)$ for companies, the
 profits of which in 2003 ($x_1$) and 2004 ($x_2$)
 exceeded $0$, $x_1 > 0$ and $x_2 > 0$.
 }
 \label{ProfitDistribution}
 \end{minipage}
 \hfill
 \begin{minipage}[htb]{0.49\textwidth}
  \epsfxsize = 1.0\textwidth
  \epsfbox{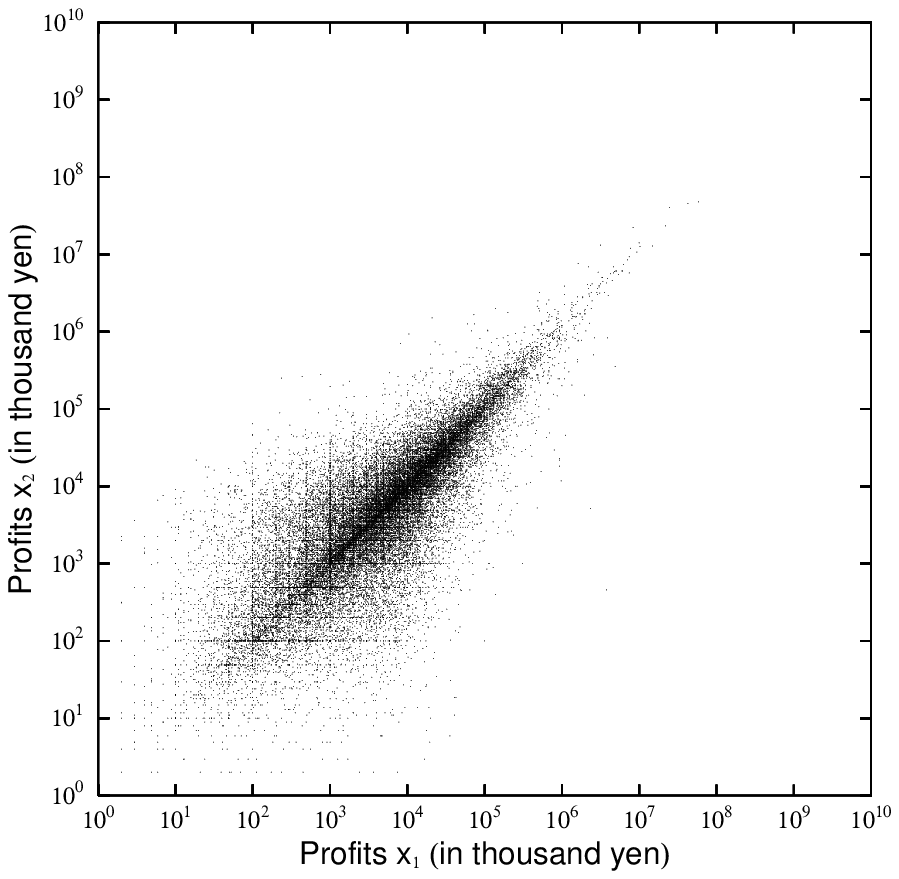}
 \caption{The scatter plot of all companies in the database, 
 the profits of which in 2003 ($x_1$) and 2004 ($x_2$)
 exceeded $0$, $x_1 > 0$ and $x_2 > 0$.
 The number of the companies is ``227,132''.}
 \label{Profit2003vsProfit2004}
 \end{minipage}
\end{figure}

In the database, 
Pareto's law (\ref{Pareto}) is observed in the large scale region
whereas it fails in the middle one (Fig.~\ref{ProfitDistribution}).
At the same time, it is confirmed that the detailed balance (\ref{Detailed balance})
holds 
(Fig.~\ref{Profit2003vsProfit2004}). 
The scatter plot in Fig.~\ref{Profit2003vsProfit2004} is different 
from one in Ref~\cite{Ishikawa2}.
The reason is that the identification of profits in 2002 and 2003 in Ref.~\cite{Ishikawa2}
was partly failed.
As a result, the pdfs of profits growth rate 
are slightly different from those in this paper.
The conclusion in Ref.~\cite{Ishikawa2} is, however, not changed.

\begin{figure}[htb]
 \begin{minipage}[htb]{0.49\textwidth}
  \epsfxsize = 1.0\textwidth
  \epsfbox{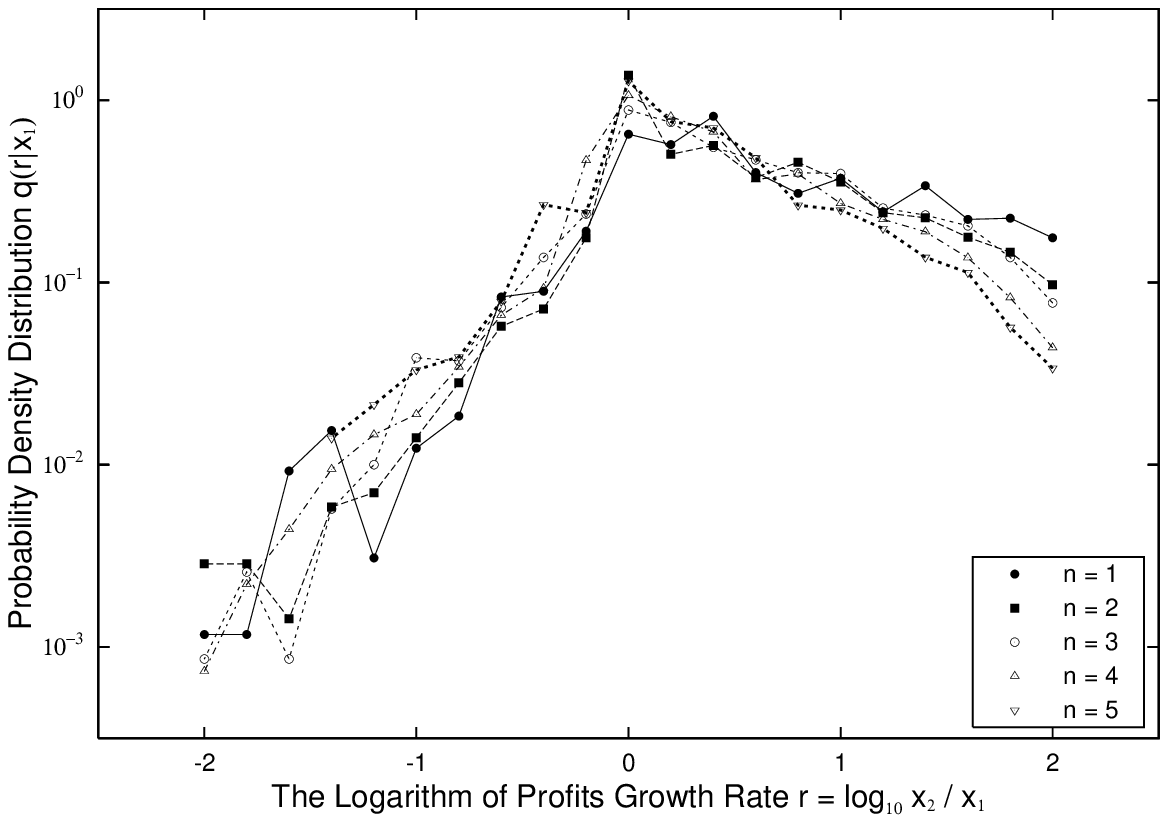}
 \end{minipage}
 \hfill
 \begin{minipage}[htb]{0.49\textwidth}
  \epsfxsize = 1.0\textwidth
  \epsfbox{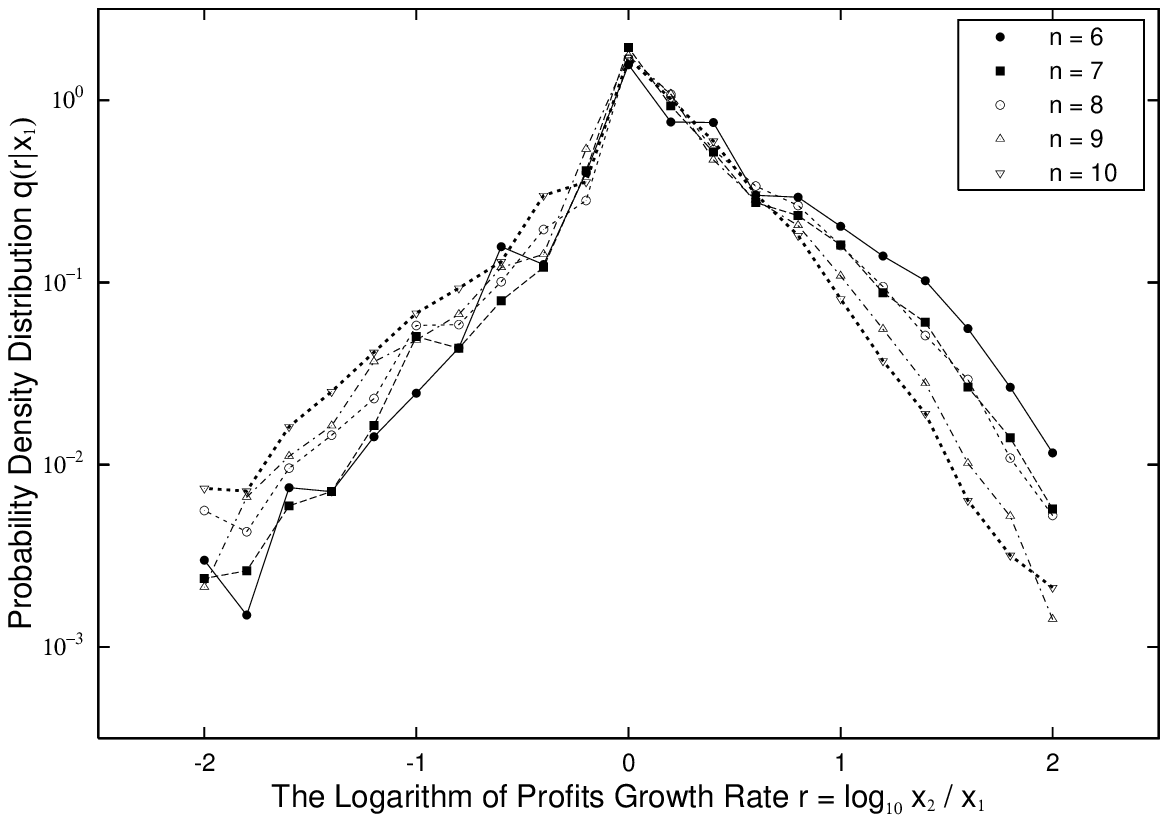}
 \end{minipage}
 \begin{minipage}[htb]{0.49\textwidth}
  \epsfxsize = 1.0\textwidth
  \epsfbox{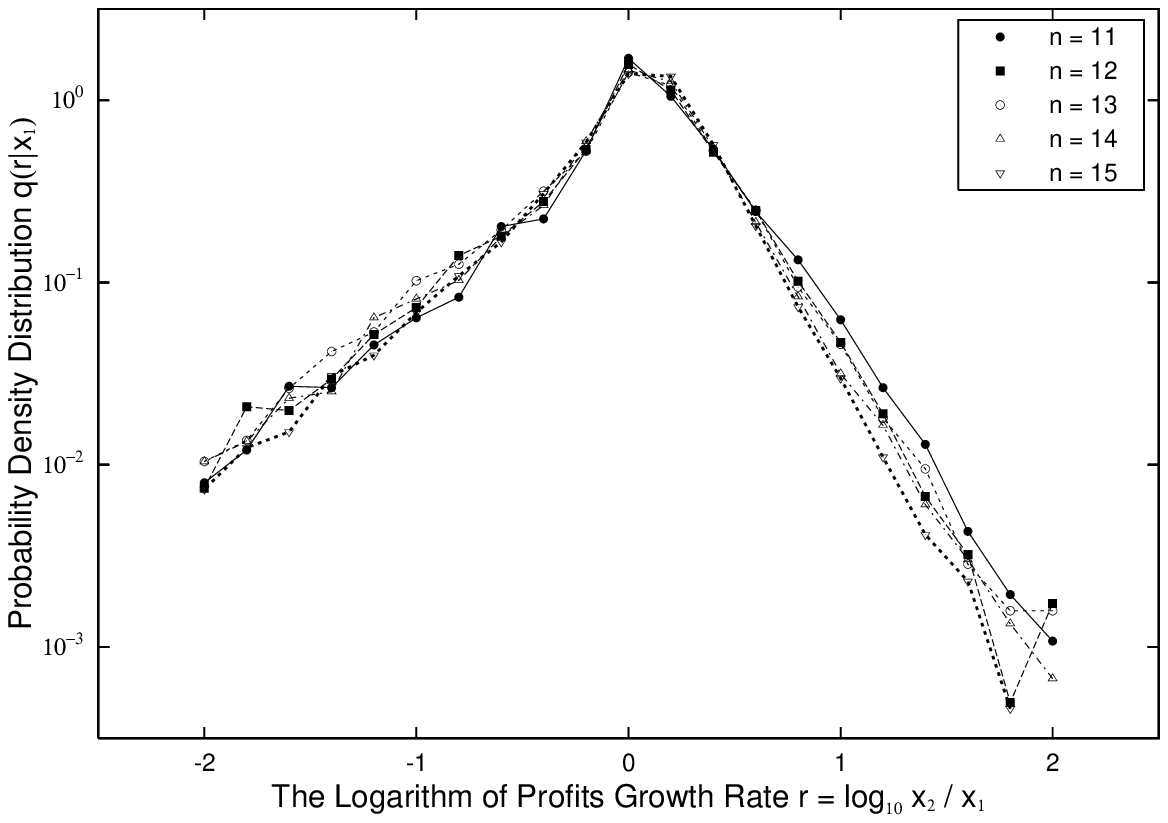}
 \end{minipage}
 \hfill
 \begin{minipage}[htb]{0.49\textwidth}
  \epsfxsize = 1.0\textwidth
  \epsfbox{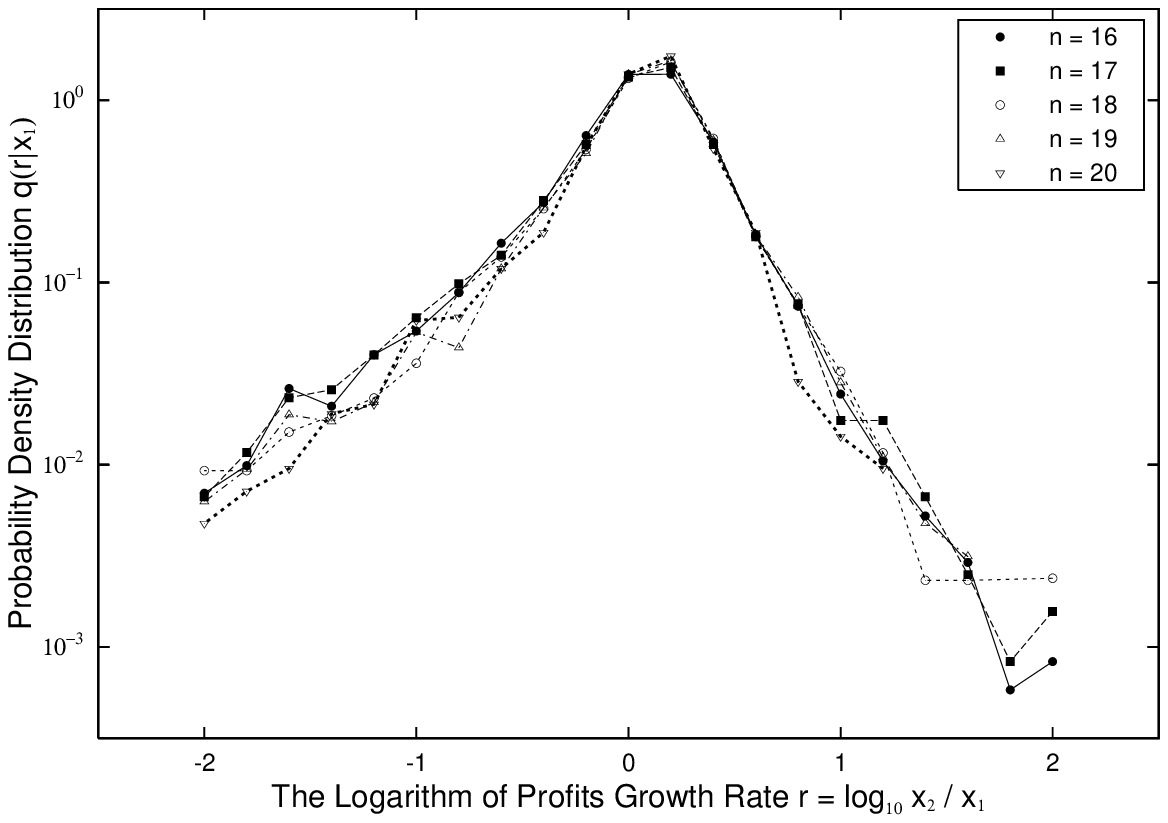}
 \end{minipage}
 \caption{The probability density distribution $q(r|x_1)$ of the log profits growth rate
 $r = \log_{10} x_2/x_1$ from 2003 to 2004.
 The data points are classified into twenty 
 bins of the initial profits with equal magnitude in logarithmic scale,
 $x_1 \in 4 \times [10^{1+0.2(n-1)},10^{1+0.2n}]~(n=1, 2, \cdots, 20)$ thousand yen.
}
 \label{ProfitGrowthRate}
\end{figure}

The breakdown of Pareto's law is thought to be caused by
the breakdown of Gibrat's law in the middle scale region.
We examine, therefore, the pdf of profits growth rate in the database.
In the analysis, we divide the range of $x_1$ into logarithmically equal bins
as $x_1 \in 4 \times [10^{1+0.2(n-1)},10^{1+0.2n}]$ thousand yen
with $n=1, 2, \cdots, 20$.
In Fig.~\ref{ProfitGrowthRate},
the probability densities for $r$ 
are expressed in the case of $n=1, \cdots, 5$,
$n=6, \cdots, 10$, 
$n=11, \cdots, 15$ and $n=16, \cdots, 20$,
respectively.
The number of the companies in Fig.~\ref{ProfitGrowthRate}
is ``$22,005$", ``$89,507$", ``$85,020$" and ``$24,203$", respectively.
Here we use the log profits growth rate $r=\log_{10} R$.
The probability density for $r$ defined by $q(r|x_1)$ is
related to that for $R$ by
    $\log_{10}Q(R|x_1)+r+\log_{10}(\ln 10)=\log_{10}q(r|x_1)$~.

From Fig.~\ref{ProfitGrowthRate},
$\log_{10}q(r|x_1)$ is approximated by linear functions of $r$ as follows
\begin{eqnarray}
    \log_{10}q(r|x_1)&=&c(x_1)-t_{+}(x_1)~r~~~~~{\rm for}~~r > 0~,
    \label{approximation1}\\
    \log_{10}q(r|x_1)&=&c(x_1)+t_{-}(x_1)~r~~~~~{\rm for}~~r < 0~.
    \label{approximation2}
\end{eqnarray}
These are expressed 
as tent-shaped exponential functions (\ref{tent-shaped1}), (\ref{tent-shaped2})
by $d(x_1)=10^{c(x_1)}/{\ln 10}$~.
In addition, the $x_1$ dependence of $c(x_1)$
is negligible for $n = 9, \cdots, 20$ (Fig.~\ref{ProfitGrowthRate}).
The validity of these approximations should be checked against the results.

We show that
the $x_1$ dependence of $t_{\pm}$ (\ref{t}) is unique
under approximations (\ref{tent-shaped1}), (\ref{tent-shaped2}).
By the use of 
the relation of
$P_{1 2}(x_1, x_2)dx_1 dx_2 = P_{1 R}(x_1, R)dx_1 dR$,
the detailed balance (\ref{Detailed balance})
is rewritten 
as 
    $P_{1 R}(x_1, R) = R^{-1} P_{1 R}(x_2, R^{-1})$~.
Substituting the joint pdf $P_{1 R}(x_1, R)$ for the conditional probability $Q(R|x_1)$,
the detailed balance is expressed as
\begin{eqnarray}
    \frac{P(x_1)}{P(x_2)} = \frac{1}{R} \frac{Q(R^{-1}|x_2)}{Q(R|x_1)}~.
\end{eqnarray}
Under approximations (\ref{tent-shaped1}) and (\ref{tent-shaped2}),
the detailed balance is reduced to
\begin{eqnarray}
    \frac{P(x_1)}{P(x_2)} = \frac{d(x_2)}{d(x_1)}~ R^{+t_{+}(x_1)-t_{-}(x_2)+1}
    \label{DE0}
\end{eqnarray}
for $R>1$.
By expanding Eq.~(\ref{DE0}) around $R=1$, the
following differential equation is obtained 
\begin{eqnarray}
    \Bigl[1+t_{+}(x)-t_{-}(x) \Bigr] \tilde{P}(x) 
        + x~ {\tilde{P}}^{'}(x) = 0,
\end{eqnarray}
where $x$ denotes $x_1$ and $\tilde{P}(x) \equiv P(x) d(x)$.
The same differential equation is obtained for $R<1$.
The solution is given by
    $\tilde{P}(x) = C x^{-1}~e^{-G(x)}$~,
where $t_+(x) - t_-(x) \equiv g(x)$ and $\int g(x)/x~dx \equiv G(x)$.

In order to make the solution 
around $R=1$
satisfies Eq.~(\ref{DE0}),
the following equation must be valid for all $R$:
\begin{eqnarray}
    -G(x)+G(R~x) = \Bigl[t_{+}(x)-t_{-}(R~x) \Bigr] \ln R~.
    \label{Kouho}
\end{eqnarray}
%
By expanding the derivative of Eq.~(\ref{Kouho}) with respect to $x$ around $R=1$, 
following differential equations are obtained
\begin{eqnarray}
    &&x~\Bigl[{t_{+}}^{''}(x)+{t_{-}}^{''}(x) \Bigr]
        +{t_{+}}^{'}(x)+{t_{-}}^{'}(x)=0~,\\
    &&2~{t_{+}}^{'}(x)+{t_{-}}^{'}(x)-3x~{t_{-}}^{''}(x)
        -x^2~\Bigl[{t_{+}}^{(3)}(x)+2~{t_{-}}^{(3)}(x) \Bigr]=0~.        
\end{eqnarray}
The solutions are given by
\begin{eqnarray}
    t_+(x) &=& -\frac{C_{-2}}{2} \ln^2 x 
        + \left(C_{+1}-C_{-1} \right) \ln x + \left( C_{+0}-C_{-0} \right)~,\\
    t_-(x) &=& \frac{C_{-2}}{2} \ln^2 x + C_{-1} \ln x + C_{-0}~.
\end{eqnarray}
To make these solutions satisfy Eq.~(\ref{Kouho}), the coefficients must be
$C_{-2}=0$ and $C_{+1}=0$.
Finally we conclude that $t_{\pm}(x)$ is uniquely expressed as 
Eq.~(\ref{t}) with $\alpha_+ = \alpha_-$.

\begin{figure}[htb]
 \begin{minipage}[htb]{0.49\textwidth}
  \epsfxsize = 1.0\textwidth
  \epsfbox{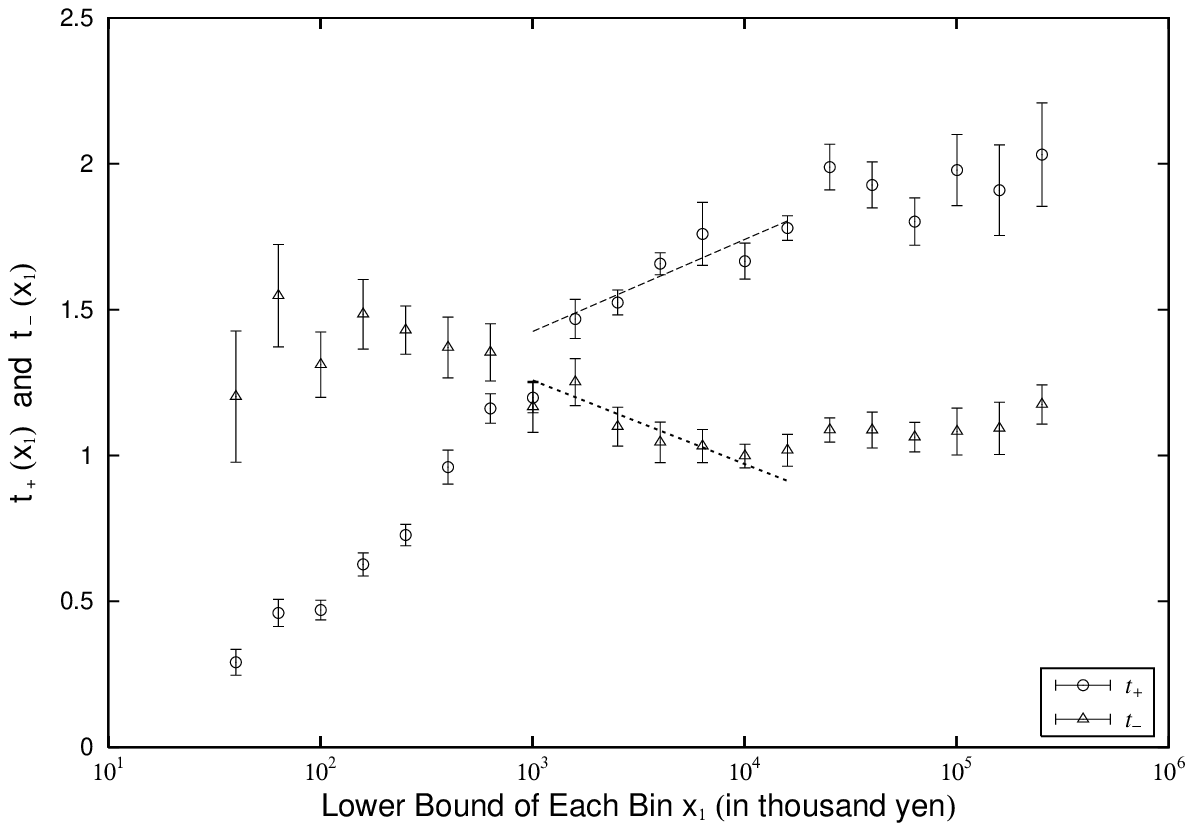}
 \caption{
  The relation between the lower bound of each bin $x_1$ and $t_{\pm}(x_1)$.
 From the left, each data point represents $n=1, 2, \cdots, 20$.}
 \label{X1vsT}
 \end{minipage}
 \hfill
 \begin{minipage}[htb]{0.49\textwidth}
  \epsfxsize = 1.0\textwidth
  \epsfbox{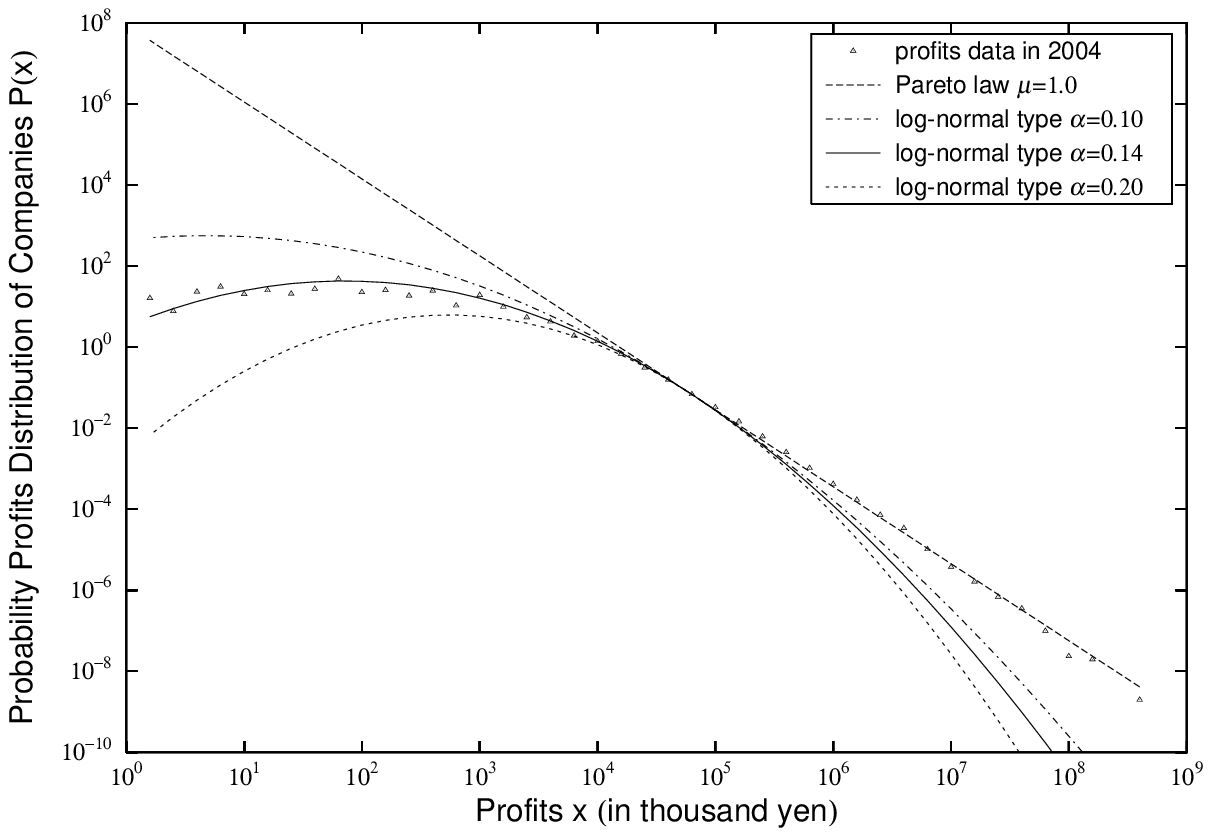}
 \caption{
  The probability distribution function (pdf) $P(x_2)$ for companies, the
 profits of which in 2003 ($x_1$) and 2004 ($x_2$)
 exceeded $0$.
}
 \label{ProfitDistributionFit}
 \end{minipage}
\end{figure}

Under approximations (\ref{tent-shaped1}) and (\ref{tent-shaped2}),
we obtain the profits pdf
\begin{eqnarray}
    \tilde{P}(x) = P(x) d(x) 
    = C x^{-\left(\mu+1\right)}~e^{-\alpha \ln^2 \frac{x}{x_0}}~,
    \label{HandM2}
\end{eqnarray}
where we use the relation (\ref{mu}) confirmed in Ref.~\cite{Ishikawa}.
In Fig.~\ref{X1vsT}, 
$t_{\pm}$ hardly responds to 
$x_1$ for $n=17, \cdots, 20$.
This means that Gibrat's law holds in the large profits region.
On the other hand,
$t_{+}$ linearly increases and $t_{-}$ linearly decreases 
symmetrically with $\log_{10} x_1$
for $n=9, 10, \cdots, 13$.
The parameters are estimated 
as Eq.~(\ref{alphaH}) and (\ref{alphaM}) with
$\alpha$ ($= \alpha_+ = \alpha_-) \sim 0.14$
and $x_0 = 4 \times 10^{1+0.2(17-1)} \sim 63,000$ thousand yen.
Because the $x_1$ dependence of $c(x_1)$
is negligible in this region, 
the profits pdf is reduced to Eq.~(\ref{HandM}).
We observe that this pdf fits with the empirical data nicely 
in Fig.~\ref{ProfitDistributionFit}.
Notice that the estimation of $\alpha$ in Fig.~\ref{X1vsT} is significant.
If we take a slightly different $\alpha$, the pdf (\ref{HandM}) 
cannot fit with the empirical data
($\alpha = 0.10$ or $\alpha = 0.20$ 
in Fig.~\ref{ProfitDistributionFit} for instance).

In this paper,
we have shown the proof that
the expression of extended Gibrat's law is unique and 
the pdf in the large and middle scale region is also uniquely derived from
the extended Gibrat's law and
the law of detailed balance.
In the proof, we have employed two approximations that
the pdf of growth rate is described as tent-shaped exponential functions
and that the value of the origin of the growth rate distribution is constant.
These approximations have been confirmed in profits data of Japanese companies
2003 and 2004.
The resultant pdf of profits has fitted with the empirical data with high accuracy.
This guarantees the validity of the approximations.

For profits data we have used, the distribution is power in the large scale region
and log-normal type in the middle one.
This does not claim that
all the distributions in the middle scale region are log-normal types.
For instance, the pdf of personal income growth rate or sales of company
is different from tent-shaped exponential functions~\cite{FSAKA}.
In this case, the extended Gibrat's law takes a different form.
In addition, we describe no pdf in the small scale region~\cite{Yakovenko}.
Because the $x_1$ dependence of $d(x_1)$ in the small scale region is not negligible
(Fig.~\ref{ProfitGrowthRate}).
Against these restrictions, the proof and the method in this paper
is significant for the investigation of distributions in the middle and small scale region.
We will report the study about these issues in the near future.
%
%


\vspace{2cm}

\end{document}